\documentclass[final]{jpp}

\usepackage[active]{srcltx}
\usepackage{graphicx}
\usepackage{natbib}

\usepackage{hhline}
\usepackage{amsmath}

\ifCUPmtlplainloaded \else
  \checkfont{eurm10}
  \iffontfound
    \IfFileExists{upmath.sty}
      {\typeout{^^JFound AMS Euler Roman fonts on the system,
                   using the 'upmath' package.^^J}%
       \usepackage{upmath}}
      {\typeout{^^JFound AMS Euler Roman fonts on the system, but you
                   dont seem to have the}%
       \typeout{'upmath' package installed. JPP.cls can take advantage
                 of these fonts, if you use 'upmath' package.^^J}%
      }
  \else
  \fi
\fi

\ifCUPmtlplainloaded \else
  \checkfont{msam10}
  \iffontfound
    \IfFileExists{amssymb.sty}
      {\typeout{^^JFound AMS Symbol fonts on the system, using the
                'amssymb' package.^^J}%
       \usepackage{amssymb}%
         \let\leq=\leqslant
         
      }{}
  \fi
\fi

\ifCUPmtlplainloaded \else
  \IfFileExists{amsbsy.sty}
    {\typeout{^^JFound the 'amsbsy' package on the system, using it.^^J}%
     \usepackage{amsbsy}}
    {}
\fi

\newsavebox{\astrutbox}
\sbox{\astrutbox}{\rule[-5pt]{0pt}{20pt}}

\newcommand{\eref}[1]{(\ref{#1})}

\title[Plane thermonuclear detonation waves initiated by proton beams]{Plane thermonuclear detonation waves initiated by proton beams and quasi-one-dimensional model of fast ignition}

\author[A. A. Charakhch'yan and K. V. Khishchenko]%
{ALEXANDER A. CHARAKHCH'YAN$^{1,2}$\break
\and KONSTANTIN V. KHISHCHENKO$^{2,3}$\thanks{Address correspondence and reprint requests to: Konstantin V. Khishchenko, Joint Institute for High Temperatures RAS, Izhorskaya~13 Bldg~2, Moscow~125412, Russia. E-mail: konst@ihed.ras.ru}}

\affiliation{$^1$Dorodnicyn Computing Centre RAS, Vavilova~40, Moscow~119333, Russia\\[\affilskip]
$^2$Moscow Institute of Physics and Technology (State University), Institutskiy~9, Dolgoprudny, Moscow Region~141700, Russia\\[\affilskip]
$^3$Joint Institute for High Temperatures RAS, Izhorskaya~13 Bldg~2, Moscow~125412, Russia}

\date{?; revised ?; accepted ?. - To be entered by editorial office}
\begin{document}

\maketitle

\newpage

{\bf Plane thermonuclear detonation waves initiated by proton beams and quasi-one-dimensional model of fast ignition}

\begin{abstract}
The one-dimensional (1D) problem on bilatiral irradiation by proton beams of the plane layer of condensed DT mixture with length $2H$ and density $\rho_0 \leqslant 100\rho_s$, where $\rho_s$ is the fuel solid-state density at atmospheric pressure and temperature of 4~K, is considered. The proton kinetic energy is 1~MeV, the beam intensity is $10^{19}$~W/cm$^2$ and duration is 50~ps. A mathematical model is based on the one-fluid two-temperature hydrodynamics with a wide-range equation of state of the fuel, electron and ion heat conduction, DT fusion reaction kinetics, self-radiation of plasma and plasma heating by $\alpha$-particles. If the ignition occurs, a plane detonation wave, which is adjacent to the front of the rarefaction wave, appears. Upon reflection of this detonation wave from the symmetry plane, the flow with the linear velocity profile along the spatial variable $x$ and with a weak dependence of the thermodynamic functions of $x$ occurs. An appropriate solution of the equations of hydrodynamics is found analytically up to an arbitrary constant, which can be chosen so that the analytical solution describes with good accuracy the numerical one. The gain with respect to the energy of neutrons $G\approx 200$ at $H\rho_0 \approx 1$~g/cm$^2$, and $G>2000$ at $H\rho_0 \approx 5$~g/cm$^2$. To evaluate the ignition energy $E_{\mathrm{ig}}$ of cylindrical targets, the quasi-1D model, limiting trajectories of $\alpha$-particles by a cylinder of a given radius, is suggested. The model reproduces the known theoretical dependence $E_{\mathrm{ig}} \sim \rho_0^{-2}$ and gives $E_{\mathrm{ig}} = 160$~kJ for $\rho_0 = 100\rho_s \approx 22$~g/cm$^3$.
\end{abstract}


\newpage
\section{Introduction}
High-gain targets for the inertial confinement fusion (ICF) are meant for the ignition in a small part of a fuel with the following propagation of the thermonuclear burn wave on its main part. As the deuterium--tritium (DT) reaction rate at the temperatures of about 10~keV is much greater than the rates of other thermonuclear reactions, the ignition of the equimolar DT mixture is considered commonly as the primary task.

There are two approaches to ICF. One of the two is based on compression of a spherical layer of the fuel by a single driver providing sufficiently high values of density and temperature \citep{Lindl-Amendt-Berger-Glendinning-Glenzer-Haan-Kauffman-Landen-Suter-2004}. At that, the thermonuclear burn wave arises in the target volume. Another approach known as the fast ignition is based on using two drivers \citep{Basov-Guskov-Feoktistov-1992, Tabak-Hammer-Glinsky-Kruer-Wilks-Woodworth-Campbell-Perry-Mason-1994, Guskov-FP-2013-eng}. The first driver compresses the target up to necessary value of density, while the second driver provides for fast rise of temperature. As a variant of such approach, one can consider ignition at the target center by a converging shock wave \citep{Shcherbakov-FP-1983-eng, Betti-Zhou-Anderson-Perkins-Theobald-Solodov-2007}. A possibility of using multi-shock wave compression for fast ignition is considered by \citet{Eliezer-MarinezVal-LPB-2011}.

The present paper is devoted to the variant of the fast ignition for which the burn wave arises near the target surface and propagates inside the target. Laser-generated beams of electrons \citep{Tabak-Hammer-Glinsky-Kruer-Wilks-Woodworth-Campbell-Perry-Mason-1994}, and ions \citep{Roth-Cowan-Key-Hatchett-Brown-Fountain-Johnson-Pennington-Snavely-Wilks-Yasuike-Ruhl-Pegoraro-Bulanov-Campbell-Perry-Powell-2001, Guskov-QE-2001-eng, Caruso-Strangio-JETP-2003, Honrubia-Fernandez-Temporal-Hegelich-Meyer-ter-Vehn-JPCS-2010} as well as heavy ion beams \citep{Churazov-Aksenov-Zabrodina-VANT-2001-eng, Medin-Churazov-Koshkarev-Sharkov-Orlov-Suslin-2002, Guskov-Ilin-Limpoukh-Klimo-Sherman-FP-2010-eng}, macroparticles \citep{Caruso-Strangio-LPB-2001} and high velocity flows of matter \citep{Guskov-FP-2013-eng, Guskov-Zmitrenko-FP-2012-eng} are considered as a driver for the fast heat of the highly-compressed fuel. Results of many theoretical works show initiation of burn waves propagating inside the target at certain values of beam parameters.

The contemporary concept of fast ignition assumes that the fuel is compressed up to the initial density $\rho_0 \sim 10^3\rho_s$, where $\rho_s$ is the density of the DT-mixture solid state at atmospheric pressure and temperature of 4~K. Such preliminary compression of the fuel is a serious technical problem due to, particularly, Rayleigh--Taylor instability \citep{Anisimov-Drake-Gauthier-Meshkov-Abarzhi-2013} arising at the stage of deceleration of a heavy shell while compressing the fuel. In the present paper, we study a possibility to use targets with the fuel density $\rho_0 = 100\rho_s$. In this connection, the work by \citet{Avrorin-Bunatyan-Gadzhiev-Mustafin-Nurbakov-Pisarev-Feoktistov-Frolov-Shibarshov-FP-1984-eng} should be mentioned, where propagation of thermonuclear detonation on unevenly compressed and heated non-spherical target is studied numerically, in particular, from a small part of the hot ($T=10$~keV) and dense ($\rho_0\approx 900\rho_s$) DT mixture to a cold and much less compressed part ($\rho_0\approx 45\rho_s$).

The first theoretical estimate of the ignition energy threshold for DT fuel of the density $\rho_0$ was obtained by \citet{Tabak-Hammer-Glinsky-Kruer-Wilks-Woodworth-Campbell-Perry-Mason-1994} in the form $E_{\mathrm{ig}}\sim \rho_0^{-2}$. The similar estimate with an improved value of the constant of proportionality was obtained by \citet{Atzeni-1999}. In many papers \citep[see, for example,][]{Caruso-Pais-1996, Atzeni-1999, Churazov-Aksenov-Zabrodina-VANT-2001-eng, Caruso-Strangio-JETP-2003}, the ignition energy threshold is determined from numerical solutions of the 2D axially symmetric problem with given radius and time dependencies of the particle beam intensity. The improved dependency $E_{\mathrm{ig}}\sim \rho_0^{-1.85}$ based on results of many computations for $\rho_0\geqslant 50$~g/cm$^3\approx 230\rho_s$ is proposed by \citet{Atzeni-1999}. Note also a number of simple burn models based on ordinary differential equations of evolution, which allow estimating of ignition parameters for uniformly heated volumes of fuel \citep{Nayak-Menon-LPB-2012}.

In the following works, shell cylindrical targets with DT pre-compressed fuel irradiated from the target end are studied using 2D numerical codes. The initiation of a thermonuclear burn wave in DT fuel of the density about $500\rho_s$ by a heavy ion beam was considered by \citet{Churazov-Aksenov-Zabrodina-VANT-2001-eng}. A possibility to decrease the ignition energy  by heating a small part of the fuel near the symmetry axis before the fuel density reaches its maximal value was considered by \citet{Caruso-Strangio-JETP-2003}. Propagation of thermonuclear burn waves within a gold shell for DT fuel density of 50, 100 and 200 g/cm$^3$, as well as the ignition of such a target by heavy ion beams were considered by \citet{Ramis-Meyer-ter-Vehn-LPB-2014}.

Symmetrically converging plane thermonuclear burn waves initiated by laser pulses for the initial density $\rho_0=\rho_s$ and $5\rho_s$ have been studied by \citet{Khishchenko-Charakhchyan-VANT-2013-eng} using the model of total absorption of laser radiation in the point with the critical density. For such kind of model, epithermal particles generated by laser-plasma interaction and heating the domain of supercritical density are ignored. A slow combustion wave arises only after at least one interaction of the shock wave reflected from the symmetry plane with the ablation front and moves in the fuel preliminary compressed and heated by several shock waves. The slow wave generates before itself a compression velocity profile that increases rapidly the fuel density. As a result, the slow combustion wave can transform into two detonation waves moving in opposite directions. After reflecting the slow or detonation wave from the symmetry plane, the intensive burning continues and increases considerably the burn-up factor of the target.

Further \citep{Charakhchyan-Khishchenko-PPCF-2013}, the study of thermonuclear burn waves at $\rho_0=\rho_s$ and $5\rho_s$ was supplemented by the waves generated by proton beams. It was shown that in spite of different ways of ignition, various models of $\alpha$-particle heat, whether the burn wave remains slow or transforms into the detonation wave, and regardless of way of such a transformation, the final value of the burn-up factor depends essentially on the only parameter $H\rho_0$, as in the known approximate formulas for the fuel expansion in spherical geometry \citep{Basko-2009-eng}.

\begin{figure}
\begin{center}
\includegraphics[width=0.5\columnwidth]{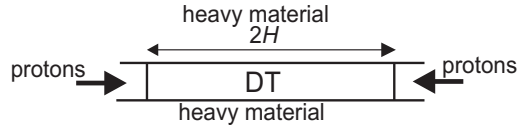}
\end{center}
\caption{The target scheme.} \label{fig1}
\end{figure}

In the present paper, we extend our study to the more wide range of the initial density $5\rho_s \leqslant \rho_0 \leqslant 100\rho_s$. The 1D problem to be considered can be treated as a rough model of burning the target shown in Figure \ref{fig1}. The fuel of the initial density $\rho_0$ is in the cylindrical channel with the length of $2H$ surrounded by a shell of a heavy material. Following, for example, \citet{Caruso-Strangio-JETP-2003}, the monoenergetic beam of protons of kinetic energy 1~MeV is considered as the ignition driver. The beam intensity is 10$^{19}$~W/cm$^2$ and the duration is 50~ps. The beam of the same energy with the intensity 10$^{18}$~W/cm$^2$ and the duration 500~ps is also considered. The burn-up factor and the gain obtained from the 1D calculation should be considered as the maximum possible values for the target of the same initial density and length.

For the first time such type of target was suggested by \citet{Pashinin-Prokhorov-JETP-1972-eng} for compression of a gaseous deuterium by laser pulses. Note that the target was experimentally studied by \citet{Stockl-Tsakiris-LPB-1991}.

The heavy shell can be compressed by a magnetic field \citep{Prut-Khrabrov-Matveev-Shibaev-1979-eng} or by heavy ions depositing energy to a certain shell of a multilayer target \citep{Churazov-Aksenov-Zabrodina-VANT-2001-eng, Dolgoleva-Zabrodin-VANT-2006-eng}. So we suppose that it is possible to create the configuration close to that shown in Figure \ref{fig1} with a cylinder of the fuel compressed to the necessary density and with two holes for injecting the proton beams.

Let us assume that the key effect determining the fuel ignition is the heat by $\alpha$-particles while the lateral expansion of the fuel, the heat transfer from the fuel to the shell and the self-radiation of plasma are insignificant. Such assumption has certain grounds. For the initial density $\rho_0 = 100\rho_s\approx 22$~g/cm$^3$, the shell density in its compression can exceed the above value more than 10 times. For example, \citet{Ramis-Meyer-ter-Vehn-LPB-2014} considered the gold shell with a density of 400~g/cm$^3$. For the time 150~ps, during which a detonation wave arises in our calculations (see section \ref{ignition}), the fuel radius in the target can be changed insignificantly. In the case of compression by a strong magnetic field, the heat flux between the fuel and the shell can be essentially diminished \citep{Pashinin-Prokhorov-JETP-1972-eng}. Finally, we refer to the work by \citet{Churazov-Aksenov-Zabrodina-VANT-2001-eng}, where the ignition of a similar target by a heavy ion beam is simulated numerically. The authors conclude that the radiative transfer is not determinative when the ignition.

If the above assumption is true, there is an interesting possibility to evaluate the ignition energy of the cylindrical target $E_{\mathrm{ig}}$ using 1D calculations. To ignite the target, it is necessary to leave a significant part of the $\alpha$-particles energy within the fuel. To determine approximately the ignition energy of the cylindrical target in Figure~\ref{fig1}, we introduce the parameter $R_{\alpha}$ and solve the 1D problem taking into account the escape of $\alpha$-particles from a cylinder of the radius $R_{\alpha}$ identifying the latter with the beam and the fuel radiuses. If the solution of the 1D problem contains the thermonuclear burn wave, we set $E_{\mathrm{ig}}=\pi R_{\alpha}^2 I(\infty)$, where $I=I(t)$ is the proton beam energy inserted at the time $t$ per unit of cross-sectional area
$$
I(t)=\int\limits_0^t J_b(t') \mathrm{d}t',
$$
where $J_b(t)$ is the beam intensity given in the 1D problem.

The trajectories of $\alpha$-particles are relying straight lines. A possibility to use strongly magnetized cylindrical shells confining the trajectories of $\alpha$-particles and therefore reducing the beam area necessary for the ignition \citep{Kemp-Basko-Meyer-ter-Vehn-2003} is not considered here. For simulation of $\alpha$-particle transport we use the track method \citep{Brueckner-Jorna-1973, Duderstadt-Moses-1982} and its modification that is a numerical method for the known Cauchy problem for the kinetic steady-state homogeneous equation in Fokker--Plank approximation \citep{Guskov-Rozanov-1982-eng}.

The argument in favor of the quasi-1D model described above is its qualitative agreement with theoretical estimates and numerical simulations of the fuel ignition by the particle beam of the radius much less than the fuel size. Apart from numerical results presented in section \ref{integ-char}, one can see such agreement from the following simple reasoning. The mean free-path length of $\alpha$-particles during the isochoric heating is about $l_\alpha \sim \rho_0^{-1}$. The energy fraction of $\alpha$-particles that remains within the channel of the radius $R_{\alpha}$ can be estimated as $\beta \approx R_{\alpha}/l_\alpha \sim R_{\alpha}\rho_0$. Assuming that $\beta$ is independent of $\rho_0$, obtain $R_{\alpha}\sim \rho_0^{-1}$, and, setting that $I(t)$ is also independent of $\rho_0$, obtain the known estimate $E_{\mathrm{ig}}\sim \rho_0^{-2}$ \citep{Tabak-Hammer-Glinsky-Kruer-Wilks-Woodworth-Campbell-Perry-Mason-1994}.

A mathematical model is presented in section \ref{problem}. The target ignition by proton beams of different intensity and the same energy, as well as generation of a detonation wave are considered in section \ref{ignition}. The detonation wave reflection from the symmetry plane is studied in section \ref{reflection}. Integrated characteristics for the initial density $5\rho_s$, $25\rho_s$ and $100\rho_s$ are discussed in section \ref{integ-char}. Conclusions are summarized in section \ref{conclusions}. Additionally, the track method for simulation of $\alpha$-particle transport and its modification using the known Cauchy problem for the kinetic steady-state homogeneous equation in the framework of Fokker--Plank approximation are described in appendix.

\section{Problem statement and numerical method}\label{problem}

We use an equation of state (EOS) of hydrogen $p=p_{\mathrm{H}}(\rho, T)$,
$\varepsilon=\varepsilon_{\mathrm{H}}(\rho, T)$ based on the wide-range semiempirical EOS model \citep{Khishchenko-LiNa-JPCS-2008}. Here $p$ is the pressure, $\varepsilon$ is the specific internal energy, $\rho$ is the density and $T$ is the temperature. 
The EOS provides for a good agreement with the Thomas--Fermi model with quantum and exchange corrections \citep{Kalitkin-1960-eng, Kalitkin-Kuzmina-1975} at high densities ($\rho \gtrsim 2$~g/cm$^3$ for H) and low temperatures, as well as gets a form of model for ideal-gases mixture of electrons and nuclei at moderate densities and high temperatures. 

At the initial point of time $t=0$, a motionless plane layer of the equimolar DT mixture occupies the domain $0 \leqslant x \leqslant H$. The one-temperature EOS of the medium is described by formulas
\begin{equation} \label{Urs1T}
p=p(\rho, T)=p_{\mathrm{H}}(A^{-1} \rho, T), \quad 
\varepsilon=\varepsilon(\rho, T)=A^{-1}\varepsilon_{\mathrm{H}}(A^{-1} \rho, T),
\end{equation}
where $A=2.5$ is the atomic weight of the mixture.

The initial density of the fuel is specified as $\rho_0 = 5\rho_s$, $25\rho_s$ and $100\rho_s$, where $\rho_s \approx 0.22$~g/cm$^3$. The initial temperature whether equals 300~K while the initial pressure $p_0$ is determined from the EOS, or is chosen (together with $p_0$) on the isentrope passing through the point ($\rho_s, p_a$), where $p_a=0.1$~MPa. Numerical results in both cases are very close to each other.

A free boundary with the pressure $p_a$ is initially at the point $x = H$. The monoenergetic beam of protons of kinetic energy 1~MeV acts upon this boundary during $\tau_{pb} = 50$~ps. With the exception of the short time interval $\delta\tau_{pb} = 0.02\tau_{pb}$, the beam has the constant intensity $J_0=10^{19}$~W/cm$^2$:
$$
J_b(t)=\begin{cases}
J_0 t/\delta \tau_{pb}, & t \leqslant \delta \tau_{pb}; \\
J_0, & \delta \tau_{pb} <t \leqslant \tau_{pb}; \\
0, & t > \tau_{pb}.
\end{cases}
$$

The less intensive beam of the same energy with $J_0=10^{18}$~W/cm$^2$ and $\tau_{pb} = 500$~ps is also considered. At the point $x = 0$, we set the symmetry condition that is equivalent to action of the identical proton beam on a symmetrical layer of the fuel.

Only the primary fusion reaction between deuteron and triton with $\alpha$-particle and neutron as the reaction products
\begin{equation} \label{ReacSin} 
\mathrm{D}+\mathrm{T} \rightarrow \alpha (\mathrm{3.5~MeV}) + \mathrm{N} (\mathrm{14~MeV})
\end{equation}
is taken into account. Neutrons are supposed to be escaped from the fuel without interaction.

In computing of all of the coefficients entering into the model, the plasma is supposed to be completely ionized.

A mathematical model is based on the equations of one-fluid two-temperature hydrodynamics. Electron and ion heat conduction, self-radiation of plasma and plasma heating by both the proton beam and $\alpha$-particles are taken into account \citep{Afanasyev-Gamaly-Rozanov-1982-eng}:
\begin{eqnarray}
&& \frac{\mathrm{d}\rho}{\mathrm{d}t} = - \rho  \frac{\partial u}{\partial x},
\label{Gydr0} \\
&& \rho \frac{\mathrm{d}u}{\mathrm{d}t} = - \frac{\partial p}{\partial x}, \label{Gydr1} \\
&& \rho \frac{\mathrm{d}\varepsilon_e}{\mathrm{d}t} = - p_e\frac{\partial u}{\partial x} + \frac{\partial}{\partial x} \varkappa_e \frac{\partial T_e}{\partial x} + \frac{3}{2}n_i k_\mathrm{B} \frac{T_i-T_e}{\tau_T}+D_e+W_e+R,
\label{EnEl} \\
&& \rho \frac{\mathrm{d}\varepsilon_i}{\mathrm{d}t} = - p_i\frac{\partial u}{\partial x} + \frac{\partial }{\partial x}\varkappa_i \frac{\partial T_i}{\partial x} + \frac{3}{2}n_i k_\mathrm{B} \frac{T_e-T_i}{\tau_T}+D_i+W_i,
\label{EnIon}
\end{eqnarray}
where $u$ is the mass velocity, $\mathrm{d}/\mathrm{d}t = \partial /\partial t + u\partial /\partial x$ is the Lagrangian derivative with respect to time, $p_e$ and $p_i$ are the electron and ion pressure, $p=p_e+p_i$ is the total pressure, $\varepsilon_e$ and $\varepsilon_i$ are the electron and ion specific internal energy, $T_e$ and $T_i$ are the electron and ion temperature, $\varkappa_e$ and $\varkappa_i$ are the electron \citep{Kalitkin-Kostomarov-2006-eng} and ion \citep{Silin-1971} heat conductivity. 
The third term in the right parts of Eqs. \eref{EnEl} and \eref{EnIon} defines the energy exchange between electrons and ions, $n_i = \rho/(A m_u)$ is the ion number density, $A$ is the atomic weight, $m_u$ is the atomic mass unit, $k_\mathrm{B}$ is the Boltzmann constant, $\tau_T$ is the temperature relaxation time \citep{Kalitkin-Kostomarov-2006-eng}. Apart from \citet{Kalitkin-Kostomarov-2006-eng} and \citet{Silin-1971}, for $\varkappa_e$, $\varkappa_i$ and $\tau_T$, we used formulas from \citet{Charakhchyan-Gryn-Khishchenko-JAMTP-2011-eng}. 
The rest terms in Eqs. \eref{EnEl} and \eref{EnIon} define the heating of electrons and ions by the proton beam ($D_e$ and $D_i$) and by $\alpha$-particles ($W_e$ and $W_i$) as well as the energy exchange between electrons and self-radiation of plasma ($R$). The radiation pressure and the momentum transfer under deceleration of $\alpha$-particles are neglected in the equation of motion \eref{Gydr1}. We also leave out of account terms describing the change of plasma composition induced by reaction \eref{ReacSin} in the equation of discontinuity \eref{Gydr0}.

EOS for electrons in Eqs.~(\ref{Gydr0})--(\ref{EnIon}) is taken in a form, which corresponds to thermal contribution of ideal electronic gas in completely ionized plasma of hydrogen isotopes (with taking into account degeneracy):
\begin{eqnarray}
&& p_e(\rho,T)=\frac{2}{3} \rho\varepsilon_e(\rho, T),
\label{eosep} \\
&& \varepsilon_e(\rho,T)=\frac{3}{2}R_A T \frac{\beta_e T}{3 \rho^{2/3} + \beta_e T},
\label{eosee}
\end{eqnarray}
where $R_A = k_\mathrm{B} N_\mathrm{A} (A m_u)^{-1}$, $N_\mathrm{A}$ is the Avogadro constant,
\begin{equation*}
\beta_e= (\pi/3)^{2/3} \frac{m_e k_\mathrm{B}}{\hbar^2} (Am_u)^{2/3},
\end{equation*}
$m_e$ is the electron mass, $\hbar$ is the Planck constant.

EOS for ions in Eqs.~(\ref{Gydr0})--(\ref{EnIon}) is taken as follows,
\begin{eqnarray}
&& p_i(\rho,T)= p_{\mathrm{H}}(A^{-1}\rho,T) - p_e(\rho,T),
\label{eosip} \\
&& \varepsilon_i(\rho,T)= A^{-1}\varepsilon_{\mathrm{H}}(A^{-1}\rho,T) - \varepsilon_e(\rho,T).
\label{eosie}
\end{eqnarray}
So, in the case of temperature equality $T=T_e=T_i$, the EOS (\ref{Urs1T}) is satisfied, $p_e+p_i=p(\rho,T)$, $\varepsilon_e+\varepsilon_i=\varepsilon(\rho,T)$.

It should be stressed that, at high temperatures, $T \gg 3\beta_e\!^{-1}\rho^{2/3}$, relations (\ref{eosep}) and (\ref{eosee}) have forms of pressure and internal energy of mono-particle ideal gas of Boltzmann,
\begin{equation*}
p_e(\rho,T)=\rho R_A T, \quad
\varepsilon_e(\rho,T)=\frac{3}{2}R_A T.
\end{equation*}
The same relations take place at high temperatures for ionic components (\ref{eosip}) and (\ref{eosie}) in the EOS model from \citet{Khishchenko-LiNa-JPCS-2008}:
\begin{equation*}
p_i(\rho,T)=\rho R_A T, \quad
\varepsilon_i(\rho,T)=\frac{3}{2}R_A T.
\end{equation*}
This implies in particular that for the problems in which the temperature difference of electrons and ions occurs at a high degree of plasma heating when the Boltzmann ideal gas approximation is applicable for all kinds of particles, and ionization is complete, EOS of hydrogen isotopes in Eqs. (\ref{Gydr0})--(\ref{EnIon}) can be taken as follows \citep{Charakhchyan-Gryn-Khishchenko-JAMTP-2011-eng},
\begin{equation}
p_e=\frac{1}{2}p(\rho,T_e), \quad \varepsilon_e=\frac{1}{2}\varepsilon(\rho,T_e), \quad p_i=\frac{1}{2}p(\rho,T_i), \quad \varepsilon_i=\frac{1}{2}\varepsilon(\rho,T_i),
\label{gryn}
\end{equation}
where functions of $p$ and $\varepsilon$ correspond to the one-temperature case (\ref{Urs1T}).
Such an approach to the separation of pressure and internal energy on the electronic and ionic parts was used previously \citep{Khishchenko-Charakhchyan-VANT-2013-eng, Charakhchyan-Gryn-Khishchenko-PESM-2013, Charakhchyan-Khishchenko-PPCF-2013} with the EOS model from \citet{Khishchenko-LiNa-JPCS-2008}. Calculations presented further with taking into account the degeneracy of electrons by Eqs. (\ref{eosep}) and (\ref{eosee}) did not show significant differences between the obtained results and the case of using Eqs. (\ref{gryn}).

The number of events of reaction \eref{ReacSin} per unit time and unit volume is as follows from \citet{Afanasyev-Gamaly-Rozanov-1982-eng}:
$$
F=n_{\mathrm{D}} n_{\mathrm{T}} \langle \sigma v \rangle_{\mathrm{DT}},
$$
where $n_{\mathrm{D}}$ and $n_{\mathrm{T}}$ are the deuteron and triton number density respectively, $\langle \sigma v \rangle_{\mathrm{DT}}$ is the ion-temperature dependence of the reaction rate averaged over Maxwellian distribution of ions \citep{Brueckner-Jorna-1973}. The burnout of fuel nuclei is described by the equation
\begin{equation} \label{n-D}
\frac{\mathrm{d}n_{j}}{\mathrm{d}t} = - n_{j}  \frac{\partial u}{\partial x} - F,
\end{equation}
where subscripts $j={}$D and T correspond to the cases of deuterium and tritium.

Our model of $\alpha$-particle heat considers the fuel burnout, but ignores the change of the density and EOS due to the change of the plasma composition. The computations for the model of local heat by $\alpha$-particles performed previously \citep{Khishchenko-Charakhchyan-VANT-2013-eng} show, that the difference between numerical results of simulations with and without taking into account the change of the plasma composition turns out insignificant.

Following \citet{Lindl-1995, Basko-2009-eng}, we introduce the local burn-up factor 
\begin{equation*} 
B_{\mathrm{loc}} = \frac{n_{\mathrm{R}}}{n_{\mathrm{R}}+n_{\mathrm{D}}},
\end{equation*}
where the number density of the fusion reaction events $n_{\mathrm{R}}$ is defined by the equation
\begin{equation} \label{n-R}
\frac{\mathrm{d}n_{\mathrm{R}}}{\mathrm{d}t} = - n_{\mathrm{R}}  \frac{\partial u}{\partial x} + F,
\end{equation}
and by the initial condition $n_{\mathrm{R}}=0$ at $t=0$. Excluding the derivative $\partial u/\partial x$ from \eref{n-D} and \eref{n-R}, one can obtain the following ordinary differential equation along trajectories of Lagrangian particles
\begin{equation} \label{dBloc-dt}
\frac{\mathrm{d} B_{\mathrm{loc}} }{\mathrm{d}t}  = \chi_B (1-B_{\mathrm{loc}}),
\end{equation}
where the function $\chi_B = F/n_D = n_{\mathrm{T}} \langle \sigma v \rangle_{\mathrm{DT}}$  is the burn-up rate. The solution of \eref{dBloc-dt} satisfying the initial condition $B_{\mathrm{loc}}=0$ at $t=0$ has the form 
\begin{equation} \label{Bloc-t}
B_{\mathrm{loc}}(s,t)=1-\exp\Biggl(-\int\limits_0^t \chi_B (s,t') \mathrm{d}t'\Biggr),
\end{equation}
where $$s(x,t)=\int\limits_0^x \rho(x', t) \mathrm{d}x'$$ is the Lagrangian coordinate.

We use two models of $\alpha$-particle heat. The first one is the track method \citep{Brueckner-Jorna-1973, Duderstadt-Moses-1982} based on simple physical reasons applying to a discrete media. The domain of $x$ is divided into cells by a numerical grid. The number of $\alpha$-particles per unit cross-section area which are produced within a grid cell is divided into several angular groups basing on the uniform directional distribution of $\alpha$-particles. The $\alpha$-particles of one angular group move along the ray intersecting one or more grid cells and are decelerated in compliance with the equation
\begin{equation} \label{particle-dec}
v \frac{\mathrm{d} v}{\mathrm{d} \xi} = a(x,v), \quad v(x_0)=v_0,
\end{equation}
where $v$ is the $\alpha$-particle velocity, $v_0 \approx 13$~Mm/s is the initial $\alpha$-particle velocity, $a(x,v)$ is the deceleration (negative acceleration) of $\alpha$-particles in plasma, $x=x_0+\xi\mu $, $x_0$ is the center position of the grid cell where the group of $\alpha$-particles is produced, $\xi$ is the coordinate along the ray, $\mu$ is the cosine of the angle between the $x$-axis and the ray direction.

The deceleration $a(x,v) = a_e(T_e(x),\rho(x),v) + a_i(T_i(x),\rho(x),v)$, where the first term in the right-hand part is related to electrons \citep{Vygovskii-Ilin-Levkovskii-Rozanov-Sherman-1990} as well as the second term---to ions \citep{Sivukhin-1964-inproc-eng}. The functions $a_e(T_e,\rho,v)$ and $a_i(T_i,\rho,v)$ from \citet{Guskov-Rozanov-1982-eng} were also used in our simulations and gave close results.

Solution of Eq. \eref{particle-dec} together with condition $v \geqslant v_{\mathrm{th}}(x)$, where $v_{\mathrm{th}} = (3k_\mathrm{B}T_i/m_\alpha){}^{1/2}$ is the velocity of $\alpha$-particle thermalization, $m_\alpha$ is the $\alpha$-particle mass, enables one to determine the contributions of the group to the right-hand parts $W_e$ and $W_i$ of Eqs. \eref{EnEl} and \eref{EnIon} for the respective grid cells (see appendix).

If plasma is located within a cylinder, the $\alpha$-particles escaping from the cylinder do not contribute to the plasma heating. The 1D track method enables us to take into account approximately the above 3D effect. We introduce a length parameter $R_\alpha$ and consider the solution of Eq. \eref{particle-dec} along the bounded interval $0\leqslant \xi\leqslant \xi_{\max}$, where 
\begin{equation} \label{quasi1D}
\xi_{\max}=R_\alpha (1-\mu^2)^{-1/2}
\end{equation}
corresponds to the intersection point of the ray and the lateral boundary of the cylinder of the radius $R_\alpha$, which symmetry axis coincides with the $x$-axis.

Let $R_\alpha \rightarrow 0$. Then $\xi_{\max} \rightarrow 0$ for all the interval $-1\leqslant \mu \leqslant 1$ with the exception of its extreme points $\mu=\pm 1$. The right parts of Eqs. \eref{EnEl} and \eref{EnIon}, which are integrals along $\mu$, $W_e\rightarrow 0$, $W_i \rightarrow 0$, and the ignition is impossible. By increasing $R_\alpha$, one can determine its value, starting from which the burn wave arises, and interpret it as the beam radius necessary for the ignition.

The second model of $\alpha$-particle heat is based on the kinetic steady-state equation for the distribution function $f(x,v,\mu)$ in Fokker--Plank approximation. If all of the produced $\alpha$-particles have the same initial velocity $v_0$, and the diffusion of the function $f(x,v,\mu)$ in the velocity space can be ignored, the known Cauchy problem for the kinetic homogeneous equation arises \citep{Guskov-Rozanov-1982-eng}:
\begin{equation} \label{ini-kin-eq}
\mu v\frac{\partial f}{\partial x} + \frac{\partial af}{\partial v} =0, \quad v_{\mathrm{th}}(x)\leqslant v \leqslant v_0, \quad f(x,v_0,\mu)=-\frac{\tilde F(x,\mu)}{a(x,v_0)},
\end{equation}
where $\tilde F(x,\mu)$ is the distribution function upon $\mu$ of the production rate of $\alpha$-particles in a unit volume near the point $x$. Since 
$$\int\limits_{-1}^1\tilde F(x,\mu)\mathrm{d}\mu = F(x),$$ 
for isotropic distribution of produced $\alpha$-particles, $\tilde F(x,\mu) = F(x)/2$.

The right-hand parts of Eqs. \eref{EnEl} and \eref{EnIon} have the form
$$
W_{e,i}(x)=-m_\alpha \int\limits_{-1}^1 \int\limits_{v_{\mathrm{th}}(x)}^{v_0} f(x,v,\mu) a_{e,i}(x,v) v\mathrm{d}v
\mathrm{d}\mu.
$$

As shown in appendix, a minor modification of the track method is a numerical method for the Cauchy problem \eref{ini-kin-eq}. The quasi-1D model is similar to that described above for the track method.

The intensity of a monoenergetic proton beam is determined by the proton velocity $v$ as $J=n_p v m_p v^2/2$, where $n_p$ is the proton number density, $m_p$ is the proton mass. Protons are supposed to be decelerated in the plasma according to Eq. \eref{particle-dec} for $\mu=-1$ and $x_0=x_b$. The value of $n_p$ is determined by given values of the initial proton velocity and the boundary beam intensity $J_b(t)$. Assuming that $n_p$ is independent of $x$, the function $J(x)$ is determined by the solution of Eq. \eref{particle-dec} $v(x)$. The right-hand parts of Eqs. \eref{EnEl} and \eref{EnIon} are
$$
D_e=\frac{a_e}{a_e+a_i}\frac{\partial J}{\partial x}, \quad D_i=\frac{a_i}{a_e+a_i}\frac{\partial J}{\partial x}.
$$

Self-radiation of plasma is described by the steady-state transfer equation in the diffusion approximation by solid angle \citep{Zeldovich-Raizer-1967}. Similarly to \citet{Marchuk-Imshennik-Basko-2009-eng}, we take into account the cooling of electrons by the inverse Compton effect using the known approximate formula \citep{Zeldovich-UFN-1975-eng, Basko-2009-eng}. Resulting equations are as follows,
\begin{equation} \label{rad-dif}
\frac{\partial q_\nu}{\partial x}=\kappa \big(B_{\nu}(T_e) - u_\nu\big), \quad 
\frac{\partial u_\nu}{\partial x}=-3\kappa q_\nu,
\end{equation}
where $\nu$ is the frequency, $B_{\nu}(T_e)$ is the Planck function, $\kappa = \kappa(\rho,T_e,\nu)$ is the absorption coefficient with accounting for the induced emission. The term in the right-hand part of Eq. \eref{EnEl} has the form
\begin{equation} \label{int-nu}
R=-\frac{\partial Q}{\partial x}-\frac{4\sigma_\mathrm{T} n_e U}{m_ec^2}k_\mathrm{B} (T_e-T_r), \end{equation} where
$$Q=\int\limits_0^{\infty} q_\nu \mathrm{d} \nu, \quad U=\int\limits_0^{\infty} u_\nu \mathrm{d} \nu,$$
$n_e=z n_i$ is the electron number density, $m_e$ is the electron mass, $c$ is the speed of light, $\sigma_\mathrm{T}$ is the Thomson scattering cross-section of photons by free electrons, $T_r$ is the photon temperature determined by the equality
$$
\int\limits_0^{\infty} B_{\nu}(T_r) \textrm{d} \nu = U.
$$

Numerical method for the simulations is based on splitting into physical processes. The Godunov first order accurate method in Lagrangian variables \citep{Godunov-1959-eng} is used for the hydrodynamics equations. For the heat conduction and the energy exchange between electrons and ions, the implicit over time method is used. The uniform Lagrangian grid contains from 350 to 700 nodes. Number of angular groups in the track method varies from 12 to 24. The results of calculations vary insignificantly when changing to double the number of grid points and angular groups.

The grid on the frequency $\nu$ for solving of Eqs. \eref{rad-dif} occupies the range from 4 to 8 decimal exponents. Number of grid nodes per a decimal exponent varies from 5 to 20. The integration with respect to the frequency in Eq. \eref{int-nu} is performed by the trapezium method.

\section{Ignition}\label{ignition}

Let us consider the initial stage of the target ignition, which is limited to the proton beam duration, $t \leqslant \tau_{pb}$. Suppose the heating by the proton beam is so fast that the motion of the fuel and its density change can be ignored. Such kind of heating will be referred to as isochoric. If the thermonuclear reaction and the plasma self-radiation are also ignored, the sum of Eqs. \eref{EnEl} and \eref{EnIon} takes the form
$$
\rho_0\frac{\partial\varepsilon(x,t)}{\partial t} = \frac{\partial (J+q_e+q_i)}{\partial x}, \quad
q_e=\varkappa_e \frac{\partial T_e}{\partial x}, \quad
q_i=\varkappa_i \frac{\partial T_i}{\partial x},
$$
$\varepsilon = \varepsilon_e + \varepsilon_i$.
Integrating this equation with respect to time $t$ from 0 to $\tau_{pb}$ and with respect to the spatial coordinate $x$ from $H-l_p$ to $H$, where $l_p$ is the mean free-path length of protons, setting the heat fluxes $q_e = q_i = 0$ at the boundaries of the integration with respect to $x$ and ignoring the initial energy of the fuel, obtain
\begin{equation} \label{Emid}
\bar \varepsilon = \frac{1}{l_p} \int\limits_{H-l_p}^H \varepsilon(x, \tau_{pb}) \mathrm{d}x = \frac{I}{\rho_0 l_p}, \quad I=\int\limits_0^{\tau_{pb}}J(t')\mathrm{d}t',
\end{equation}
where $\bar\varepsilon$ is the average specific internal energy of plasma in the heating region at $t = \tau_{pb}$, $I$ is the beam energy per unit cross-sectional area. Since the dependence of the proton deceleration in plasma $a$ on the density is close to the linear one, the free-path length of protons $l_p \sim \rho_0^{-1}$. Therefore the averaged internal energy $\bar\varepsilon$ is independent of $\rho_0$ and is determined by only the beam energy $I$ and by the initial proton energy (1~MeV) which determines the free-path length.

Epithermal protons give up the greater part of their energy to electrons, while ions are heated by both epithermal protons and more hot electrons. For the given beam energy $I$, the ion temperature at $t = \tau_{pb}$ increases both with increasing $\tau_{pb}$ and with increasing $\rho_0$ because the relaxation time of the electron and ion temperatures $\tau_T \sim \rho^{-1}$. If the electron temperature at $t = \tau_{pb}$ is much greater than the ion temperature, the latter can be sufficient for the ignition after stopping the action of the proton beam at $t > \tau_{pb}$.

The isochoric heating violation is connected with the rarefaction wave moving at the speed of sound from the free boundary into the target, and with the shock wave, arising due to the rapid growth of pressure in the heating region. As the condition of the isochoric heating violation, we take the inequality $c\tau_{pb} > l_p$, where $c$ is the speed of sound. Since $l_p \sim \rho_0^{-1}$, with increasing $\rho_0$ one should decrease the beam duration $\tau_{pb}$ to keep close to the isochoric heating, which, as it follows from Eq. \eref{Emid}, needs the appropriate increase in the beam intensity $J_0$ to conserve the average internal energy in the heating region.

\begin{figure}
\begin{center}
\includegraphics[width=0.9\columnwidth]{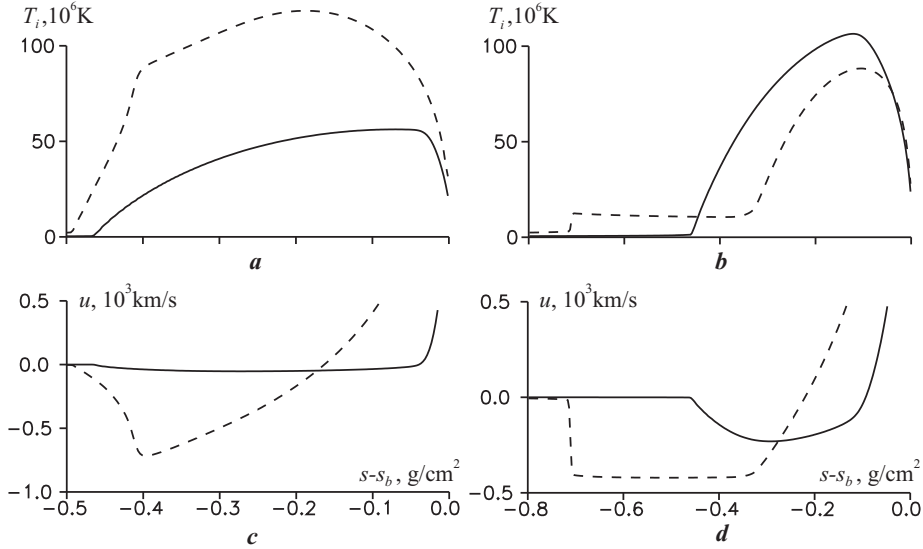}
\end{center}
\caption{
The ion temperature (\textit{a}, \textit{b}) and the mass velocity (\textit{c}, \textit{d}) as functions of the Lagrangian variable with respect to the free boundary $s-s_b$ at $t = \tau_{pb}$: $\rho_0 = 25\rho_s$, $R_{\alpha} = 0.4$~mm (\textit{a}, \textit{c}) and $\rho_0 = 100\rho_s$, $R_{\alpha} = 0.1$~mm (\textit{b}, \textit{d}) for different beams of the same energy, $J_0 = 10^{19}$~W/cm$^2$, $\tau_{pb} = 50$~ps (solid lines) and $J_0 = 10^{18}$~W/cm$^2$, $\tau_{pb} = 500$~ps (dashed lines).
} \label{fig2}
\end{figure}

The ion temperature and the mass velocity profiles by the Lagrangian variable with respect to the free boundary $s-s_b$ at the time $t = \tau_{pb}$ for the beams of the same energy and different durations $\tau_{pb} = 50$ and 100~ps, as well as for the two values of the initial density $\rho_0 = 25\rho_s$ and $100\rho_s$ are presented in Figure~\ref{fig2}. The parameter $R_{\alpha}$ limiting the trajectory of $\alpha$-particles is selected near the ignition boundary of the target by at least one of the beams ($R_{\alpha} = 0.4$ and 0.1~mm for $\rho_0 = 25\rho_s$ and $100\rho_s$ respectively).

First, we examine the case of $\rho_0 = 25\rho_s$ (see Figures \ref{fig2}\textit{a} and \ref{fig2}\textit{c}).

At $\tau_{pb} = 50$~ps, as follows from the corresponding velocity profile (the solid line in Figure~\ref{fig2}\textit{c}), the heating is almost isochoric: the formation of the shock wave has not started yet, and the rarefaction wave occupies only a small part of the heating region. Although the ion temperature (the solid line in Figure~\ref{fig2}\textit{a}) is relatively small, its subsequent growth due to heating by more hot electrons leads to the target ignition.

With increasing the duration $\tau_{pb}$ up to 500~ps, the proton beam heating is no longer isochoric. As seen in the velocity profile (the dashed line in Figure~\ref{fig2}\textit{c}), a continuous compression wave is formed inside the heating region. Then this wave should transform into a shock wave. A rarefaction wave adjoins the compression one, which in the absence of heat by $\alpha$-particles should lead to a rapid drop in the amplitude of the shock wave. However, the ion temperature (dashed line in Figure~\ref{fig2}\textit{a}) is sufficient to ignite the target. Analysis of the calculation results shows that a significant heating of the plasma by $\alpha$-particles takes place in the field of the compression wave in Figure~\ref{fig2}\textit{c}. As a result, the compression wave is rapidly converted to the detonation wave of the well-known type \citep{Landau-Lifshitz-VI-1987-eng} with a rarefaction wave, which is adjacent to the detonation wave front.

We now turn to the case $\rho_0 = 100\rho_s $ (see Figures~\ref{fig2}\textit{b} and \ref{fig2}\textit{d}).

When $\tau_{pb} = 50 $~ps, the ion temperature (the solid line in Figure~\ref{fig2}\textit{b}) is much greater than in the case of the same beam and $\rho_0 = 25\rho_s $ due to the decrease of the temperature relaxation time $\tau_T \sim \rho^{-1}$. As can be seen from the velocity profile (the solid line in Figure~\ref{fig2}\textit{d}), the shock wave begins to form, and the rarefaction wave takes until a small portion of the heating region. The ignition mechanism for these parameters will be discussed below.

For the beam with $\tau_{pb} = 500$~ps, the target does not ignite. As can be seen from the velocity profile (dashed line in Figure~\ref{fig2}\textit{d}), the shock wave at $t = \tau_{pb}$ is too far away from the heating region and no longer can transform into a detonation wave. After some time, the rarefaction wave overtakes the shock front and starts to reduce its amplitude. Calculations show that for the fuel density $\rho_0 = 10^3\rho_s $, a similar flow pattern without the target ignition occurs for the beam with $\tau_{pb} = 50 $~ps.

The aforementioned lack of the ignition by the beam with the intensity of 10$^{18}$~W/cm$^2$ for the fuel density $\rho_0 = 100\rho_s \approx 22$~g/cm$^3$, and the ignition by the beam with the intensity of 10$^{19}$~W/cm$^2$ for the same density, as well as the lack of the ignition by this beam for the fuel density $\rho_0 \approx 220$~g/cm$^3$, correspond to the density dependence of the minimum beam intensity required for ignition \citep{Atzeni-1999} based on results of two-dimensional calculations, which gives a value of approximately $6\times 10^{18}$~W/cm$^2$ for $\rho_0 = 22$~g/cm$^3$ and about $5\times 10^{19}$~W/cm$^2$ for $\rho_0 = 220$~g/cm$^3$.

Results of calculations for a target with $\rho_0 = 100\rho_s$, $R_{\alpha} = 0.1 $~mm, and for the beam with $\tau_{pb} = 50$~ps, $J_0 = 10^{19}$~W/cm$^2$ are presented below. Consider first the case of the fuel layer half-width $H=0.5$~mm, which corresponds to the value of the parameter $H\rho_0 \approx 1$~g/cm$^2$.

\begin{figure}
\begin{center}
\includegraphics[width=0.9\columnwidth]{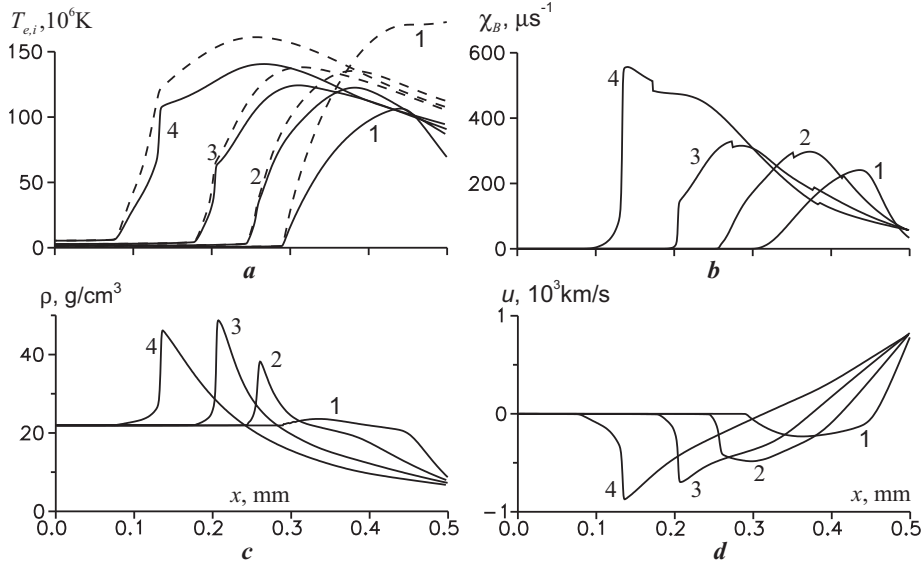}
\end{center}
\caption{
The temperature (\textit{a}, the solid lines correspond to ions, the dashed lines---to electrons), the burn-up rate (\textit{b}), the density (\textit{c}) and the mass velocity (\textit{d}) spatial profiles in the formation of the detonation wave at $t=50$ (1), 100 (2), 150 (3) and 200~ps (4) for the target with $\rho_0 = 100\rho_s$, $R_{\alpha} = 0.1$~mm, $H = 0.5$~mm and for the beam with $\tau_{pb} = 50$~ps.
} \label{fig3}
\end{figure}

Figure~\ref{fig3} shows the ignition process of the target. The functions $T_e(x) $, $T_i(x) $, $\rho (x)$, $u(x)$ and the burn-up rate $\chi_B(x)$, which determines the local burn-up factor $B_{\mathrm{loc}}$ by Eq. \eref{Bloc-t}, are given at the termination of the beam 50~ps and three subsequent times 100, 150 and 200~ps. A small discontinuity in the function $\chi_B(x)$ is caused by the discontinuity in the approximation formula for $\langle \sigma v \rangle_{\mathrm{DT}}(T_i)$ from \citet{Brueckner-Jorna-1973}.

At $t = 50$~ps, the maximal ion temperature is about 100~MK, which is sufficient for the ignition. The shock wave has not yet formed, and the burning wave, which can be identified with behavior of the function $\chi_B(x)$, is continuous. At $ t = 100$~ps, the shock wave is formed, and the burning wave is a little behind it and still continuous. At $ t = 150$~ps, the burning wave has caught up with the shock wave and transformed it into the detonation one. Comparing functions $T_i(x)$ and $\chi_B(x)$ at $t=150$ and 200~ps, one can see that the temperature and the burn-up rate at the detonation wave front quickly grow with time. Before the detonation wave one can see a precursor caused by electron heat conduction and $\alpha$-particles (near the wave front) and by the self-radiation of high-temperature plasma (at large distances from the front).

\begin{table}
\caption{
The burn-up factors for the targets with $\rho_0 = 100\rho_s$, $H = 0.5$ ($B_0$) and 2.5~mm ($B_1$), for different values of the parameter  $R_{\alpha}$ and different models of $\alpha$-particle transport: the Fokker--Plank equation (FP) and the track method (TM).
} \label{table0}
\begin{center}
\begin{tabular}{ccccc}
\hline
\multicolumn{1}{c}{$R_{\alpha}$,} & \multicolumn{2}{c}{FP} & \multicolumn{2}{c}{TM} \\
mm & $B_0$ & $B_1$ & $B_0$ & $B_1$  \\
\hline
0.15 & 0.36& 0.73 & 0.36 & 0.73  \\
0.1 & 0.32& 0.72 & 0.32 & 0.71  \\
0.05 & 0.24& 0.67 & 0.25 & 0  \\
0.03 & 0.17& 0 & 0.05 & 0  \\
0.01 & 0 & 0 & 0 & 0  \\
\hline
\end{tabular}
\end{center}
\end{table}

To conclude this section, we present the numerical results for the target with the initial density $\rho_0 = 100\rho_s$ at different values of $R_{\alpha}$, those limiting trajectory of $\alpha $-particles, and for different models of $\alpha $-particle transport: the Fokker--Plank equation and the track method. Table \ref{table0} shows the values of the burn-up factor for targets with two values of $H = 0.5$ ($B_0$) and 2.5~mm ($B_1$). These values correspond to a sufficiently late point in time, when their significant growth ceases. Zero values $B_0$ and $B_1$ means that their value is less than 0.01, and are interpreted as a lack of ignition.

At $R_{\alpha} = 0.15$~mm, the factors $B_0$ and $B_1$ are the same for both models, and slightly differ from respective values at $R_{\alpha} = 0.1$~mm. At $R_{\alpha} = 0.05$~mm, the Fokker--Plank equation still leads to the ignition of the target with somewhat smaller values of $B_0$ and $B_1$ compared with the case of $R_{\alpha} = 0.1$~mm, and in the case of the track method $B_1 = 0$. This means that the detonation wave is not formed for $H=2.5$~mm since the shock wave at large distances is extinguished by the rarefaction wave. If the symmetry plane is located close enough to the edge of the target ($H = 0.5$~mm), the shock wave at the moment of reflection is strong enough to ignite the target, creating a reflected detonation wave. At $R_{\alpha} = 0.03$~mm, the ignition is possible only if $H = 0.5$~mm with small values of $B_0$. At $R_{\alpha} = 0.01$~mm the ignition is absent. To estimate the energy of ignition, we set $R_{\alpha} = 0.1$~mm, wherein the reliable ignition occurs in both models.

\section{Reflection of detonation wave}\label{reflection}

\begin{figure}
\begin{center}
\includegraphics[width=0.9\columnwidth]{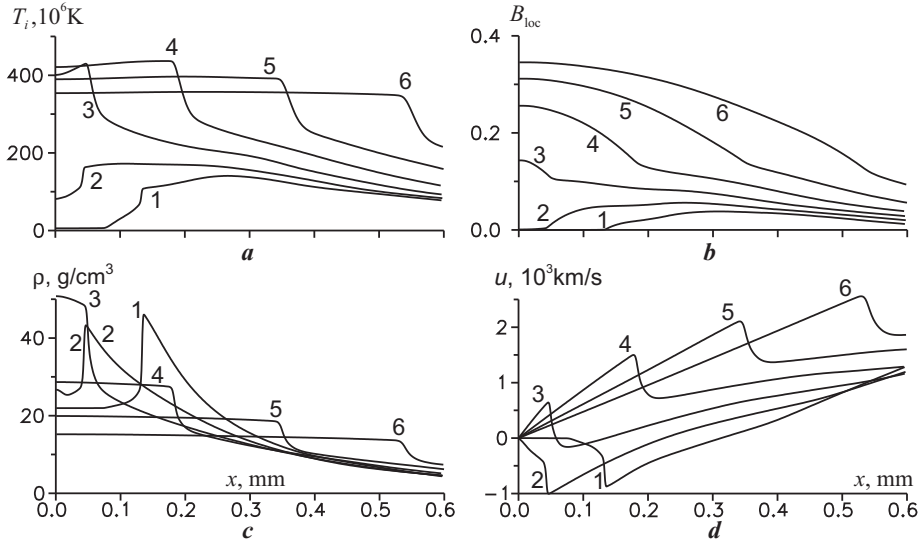}
\end{center}
\caption{
The spatial profiles of the ion temperature (\textit{a}), the local burn-up factor (\textit{b}), the density (\textit{c}) and the mass velocity (\textit{d}) in reflection of the detonation wave in points of time 200 (1), 250 (2), 300 (3), 350 (4), 400 (5) and 450~ps (6) for the parameters of the problem, as shown in Figure~\ref{fig3}.
}
\label{fig4}
\end{figure}

In the case of $H = 0.5$~mm, reflection of the detonation wave from the symmetry plane $x = 0$ is shown in Figure~\ref{fig4}. The functions $T_i(x)$, $\rho(x)$, $u(x)$ and the local burn-up factor $B_{\mathrm {loc}}(x)$ are presented in the six time points. The first two points (200 and 250~ps) meet the detonation wave before its reflection from the symmetry plane. Note the relatively small values of the local burn-up factor (about 0.05). Pay attention to the velocity profile at $t = 250$~ps (curve 2 in Figure~\ref{fig4}\textit{d}). One can see that the precursor of the detonation wave transforms into a flow with a linear velocity profile between the wave and the symmetry plane. Such flows, which are characterized by the dependence
\begin{equation} \label{LinProf}
u(x,t)=\varphi(t)x,
\end{equation}
are well known in hydrodynamics \citep{Sedov-1972} and combined into a large family of solutions, each member of that is determined by an arbitrary function of one argument and three arbitrary constants. For all of these solutions, the following formula takes place
\begin{equation} \label{LinProfRo}
\rho(s,t) = \rho(s,t_1) \exp \Biggl(-\int\limits_{t_1}^t \varphi(t') \textrm{d} t' \Biggr),
\end{equation}
which is obtained by integrating an ordinary differential equation arising under the substitution \eref{LinProf} to the equation of discontinuity \eref{Gydr0}. In the case under consideration, since $\varphi(t) < 0$, the density in the region of the precursor and, in particular, in the symmetry plane, grows with time.

The last four points of time in Figure~\ref{fig4} (300, 350, 400 and 450~ps) give the flow pattern after reflection of the detonation wave. One can see that the flow with a linear velocity profile arises again between the reflected detonation wave and the symmetry plane. Its special feature is the proximity of the thermodynamic functions to the constant upon $x$ value, which depends on time. The local burn-up factor increases significantly after reflection of the detonation wave and continues to grow because the burn-up rate $\chi_B(x,t)$ decreases with time slowly enough, remaining almost constant function of $x$. Note that the spatial homogeneity of the thermodynamic functions after the reflected wave front is not typical for the solutions of the equations of hydrodynamics in the case of spherical or cylindrical geometry. For example, a well-known solution on the problem of a converging spherical shock wave \citep{Guderley-1942, Stanyukovich-1955-eng}, extended through time after the shock wave collapse, has zero density at the point of symmetry \citep{Goldman-1973}.

Solution of the equations of hydrodynamics with a linear velocity profile and constant on $x$ values of thermodynamic functions is contained in the above-mentioned family from \citet{Sedov-1972}. It also can be obtained directly from the equation of motion \eref{Gydr1}. Putting in Eq. \eref {Gydr1} $\partial p / \partial x = 0$ and substituting the velocity in the form \eref{LinProf}, we obtain an ordinary differential equation $\dot\varphi + \varphi^2 = 0$, whose solution has the form 
\begin{equation} \label{LinProf1}
\varphi(t)=\frac{1}{C+t},
\end{equation}
where $C$ is an arbitrary constant. Substituting \eref{LinProf1} to Eq. \eref{LinProfRo}, we obtain
\begin{equation} \label{LinProfRo1}
\rho(t) = \frac{\rho_0 C}{C+t} = \rho_0 C\varphi(t),
\end{equation}
where constant $\rho_0$ is chosen so that $\rho(0) = \rho_0$.

\begin{figure}
\begin{center}
\includegraphics[width=0.45\columnwidth]{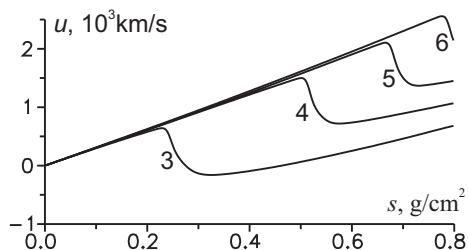}
\end{center}
\caption{
Mass velocity profiles along the Lagrangian coordinate after reflection of the detonation wave for the same parameters of the problem and the time points as in Figure~\ref{fig4}.
}
\label{fig5}
\end{figure}

Since the density $\rho$ is independent of the spatial coordinate $x$, the velocity has a linear profile along the Lagrangian coordinate $s= \rho x$, the slope of which, as follows from Eqs. \eref{LinProf}, \eref{LinProf1} and \eref{LinProfRo1}, $u/s = (\rho_0 C)^{-1}$, is independent of time. Shown in Figure~\ref{fig5} velocity profiles along the Lagrangian coordinate indicate that this property is satisfied with good accuracy for the problem in the region between the symmetry plane and the reflected detonation wave.

Using the simulation results, one can determine the approximate value of constant $C$ in Eq. \eref{LinProf1}. It is convenient to substitute time $\bar t = t - t_*$ for $t$ in Eqs. \eref{LinProf1} and \eref{LinProfRo1}, where $t_*$ is the time point of appearance of the reflected detonation wave. Note that the density on the symmetry plane as a function of time $\rho(0,t)$ increases as the incident detonation wave approaches to the symmetry plane, and begins to decrease in accordance with Eq. \eref{LinProfRo1} after appearance of the reflected wave. It is therefore natural to define the time of formation of the reflected wave $t_*$ and constant $\rho_0$ in Eq. \eref{LinProfRo1} by the condition $\max_t(\rho(0,t))=\rho(0,t_*)=\rho_0$. The result is $\rho_0 \approx 80$~g/cm$^3$, $t_* \approx 270$~ps.

The calculation results are compared with Eqs. \eref{LinProf1} and \eref{LinProfRo1} for the four time points shown in Figure~\ref{fig4} after reflection of the detonation wave, $t = 300$, 350, 400 and 450~ps. As the slope $\varphi$ and density $\rho$ for each time, the values at the point $x$ approximately two times less than the coordinate of the front of the reflected detonation wave are selected. We denote by $\bar\varphi$ and $\bar\rho $ the corresponding values included in Eqs. \eref{LinProf1} and \eref{LinProfRo1}. At first, we put $\bar\varphi = \varphi$ and calculate $C$ from Eq. \eref{LinProf1}, then $\bar\rho$ by Eq. \eref{LinProfRo1} and the relative error of calculating the density $\delta\rho = |1 - \bar\rho/\rho |$. The results of calculations are shown in the first three columns of Table~\ref{table1}. For the first three values of $t$, constant $C$ varies within 10\% and the error $\delta\rho$ is within 6\%. At the last time point $t = 450$~ps, constant $C$ changes by about 10\% more, and the error $\delta\rho$ grows to 14\%. Nevertheless, it is possible to select a single value of constant $C$ for all four points of time and get a relatively small magnitude errors $\delta\varphi = |1 - \bar\varphi/\varphi |$ and $\delta\rho$. Such a possibility is demonstrated in the last two columns of Table~\ref{table1}, where the values of these errors are presented for $C = 40$~ps. It can be seen that both errors are within 3\%.

\begin{table}
\caption{
Comparison of simulated values of $\varphi$ and $\rho$ with the analytical Eqs. \eref {LinProf1} and \eref{LinProfRo1}: the value of $C$ from Eq. \eref{LinProf1} as well as the relative error $\delta\rho $ for the given value $\varphi$ and the relative errors $\delta\rho_{40}$ and $\delta\varphi_{40}$ for $C = 40$~ps at some points of time $t$.
} \label{table1}
\begin{center}
\begin{tabular}{ccccc}
\hline $t$, ps & $C$, ps & $\delta\rho$ & $\delta\rho_{40}$ & $\delta\varphi_{40}$ \\
\hline
300 & 41.6 & 0.02 & 0.03 & 0.02  \\
350 & 39.8 & 0.03 & 0.02 & 0.001  \\
400 & 37.6 & 0.06 & 0.02 & 0.01  \\
450 & 33.2 & 0.14 & 0.003 & 0.03  \\
\hline
\end{tabular}
\end{center}
\end{table}

\begin{figure}
\begin{center}
\includegraphics[width=0.9\columnwidth]{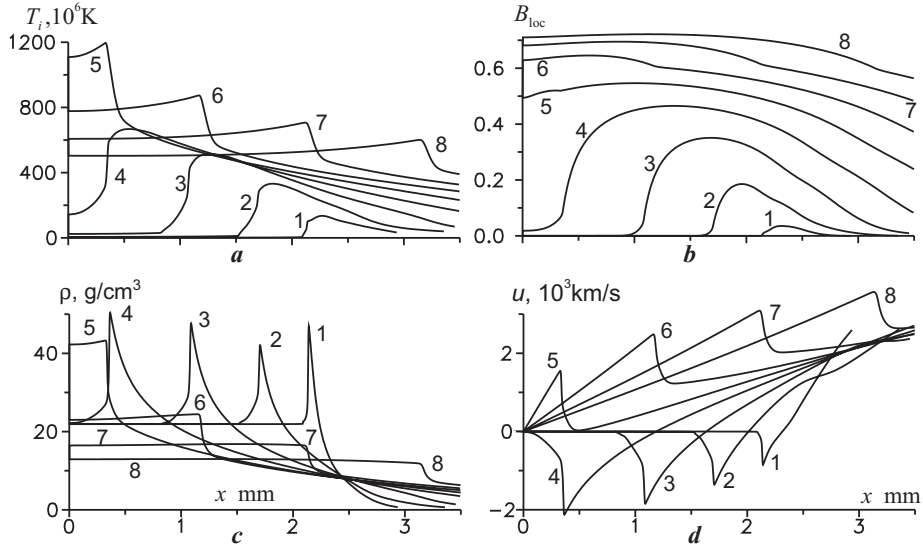}
\end{center}
\caption{
Spatial profiles of the ion temperature (\textit{a}), the local burn-up factor (\textit{b}), the density (\textit{c}) an the mass velocity (\textit{d}) at the time points 0.2 (1), 0.4 (2), 0.6 (3), 0.8 (4), 1 (5), 1.2 (6), 1.4 (7) and 1.8~ns (8), corresponding to the incident (1--4) and the reflected (5--8) detonation waves for $H = 2.5$~mm (another parameters of the problem are the same as in Figure~\ref{fig3}).
} \label{fig6}
\end{figure}

Let us consider now the target with the layer half-thickness $H = 2.5$~mm, which corresponds to the value of the parameter $H\rho_0 \approx 5$~g/cm$^2$. Figure~\ref{fig6} shows the ion temperature, the density, the mass velocity and the local burn-up factor as functions of $x$ at eight points of time. The first four points give a picture of the flow before the detonation wave reflection from the symmetry plane. Detonation wave intensity increases with time, as can be seen from the profiles of the ion temperature and the mass velocity, reaching values of the order of 700~MK and 2~Mm/s at the instant of reflection. Such a high velocity in the incident detonation wave at the reflection leads to the ion temperature after reflection of about 1.3~GK (see curve 5 in Figure~\ref{fig6}\textit{a}). Further, as in the above case $H = 0.5$~mm, the flow with a linear velocity profile and the weak dependence of the thermodynamic functions upon $x$ arises between the reflected detonation wave and the symmetry plane.

Note the rapid rise of the local burn-up factor after reflection of the detonation wave as close to the symmetry plane as in the periphery. At the last instant shown in Figure~\ref{fig6}, the local burn-up factor in the region $0 \leqslant x \leqslant 3.5$, containing about 80\% of the fuel mass, varies from about 0.7 to 0.6.

\begin{figure}
\begin{center}
\includegraphics[width=0.25\columnwidth]{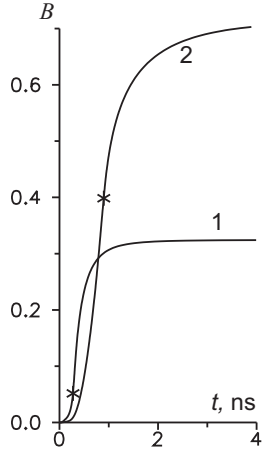}
\end{center}
\caption{
The time dependence of the burn-up factor for $H=0.5$ (1) and 2.5~mm (2) (another parameters of the problem are the same as in Figure~\ref{fig3}); asterisks denote the instant of the detonation wave reflection.
} \label{fig7}
\end{figure}

The burn-up factor $B(t)$ is presented in Figure~\ref{fig7} for the two above values of $H$. The point of reflection of the detonation wave $t = t_*$ is indicated. One can see that the expansion stage $t>t_*$ gives the main contribution to the final value of the factor in the case of $H=0.5$~mm. For $H=2.5$~mm the burn-up factor increases from 0.4 to 0.7 at the expansion stage.

\section{Integrated characteristics}\label{integ-char}

\begin{table}
\caption{
The ignition energy for one proton beam $E_{\mathrm{ig}}$, the target mass $M$, the burn-up factor $B$ and the gain $G$ for three values of $\rho_0$, the cylinder radius $R_{\alpha} \sim \rho_0^{-1}$ and different values of the layer half-width $H$ determined by two given values of the parameter $H\rho_0$.
} \label{table2}
\begin{center}
\begin{tabular}{ccccccccccc}
\hline
$\rho_0/\rho_s$ & $R_{\alpha}$, & $E_{\mathrm{ig}}$, & \multicolumn{4}{c}{$H\rho_0 \approx 1$~g/cm$^2$} & \multicolumn{4}{c}{$H\rho_0 \approx 5$~g/cm$^2$} \\
& mm & MJ & $H$, mm & $M$, mg & $B$ & $G$ & $H$, mm & $M$, mg & $B$ & $G$ \\
\hline
5 & 2 & 62 & 10 & 275 & 0.34 & 200 & 50 & 1400 & 0.68 & 2000 \\
25 & 0.4 & 2.5 & 2 & 11 & 0.33 & 200 & 10 & 55 & 0.68 & 2000 \\
100 & 0.1 & 0.16 & 0.5 & 0.7 & 0.34 & 200 & 2.5 & 3.4 & 0.73 & 2200 \\
\hline
\end{tabular}
\end{center}
\end{table}

Generalized characteristics of the target for three values of the initial density $\rho_0 = 5 \rho_s$, $25\rho_s$ and $100\rho_s$ are shown in Table~\ref{table2}. For $\rho_0 = 100\rho_s $ the characteristics are obtained by the results of the above calculations with the parameter $R_{\alpha} = 0.1$~mm. For other values of $\rho_0$ the parameter $R_{\alpha}$ is chosen from the condition $\rho_0 R_{\alpha} = \mathrm{const}$. For each value of $\rho_0$ two values of the layer half-width $H$, corresponding to the two given values of the parameter $H\rho_0 \approx 1$ and 5~g/cm$^2$, are chosen. Interpreting the parameter $R_{\alpha}$ as the radius of the cylindrical target, shown in Figure~\ref{fig1}, one can define the target ignition energy $E_{\mathrm{ig}} = \pi R_{\alpha}^2 I(\infty)$ and the mass $M = 2H \pi R_{\alpha}^2\rho_0$. These values are given in the table together with the values of the gain with respect to the energy of neutrons and the burn-up factor after the burning process.

As in the case of the well-known approximate formula for the expansion of a spherical target \citep{Basko-2009-eng}, the burn-up factor $B$, as well as the gain $G$, are determined by the parameter $H \rho_0$. If $H\rho_0\approx 1$~g/cm$^2$, $B\approx 0.3$, $G \approx 200$. If $H\rho_0 \approx 5$~g/cm$^2$, the burn-up factor increases by more than a factor of 2, which, together with the increase in the fuel mass 5 times, gives the gain $G>2000$.

Since the parameter $R_{\alpha}$, determined by the condition $R_{\alpha} \rho_0 = \mathrm{const} $, provides the target ignition, the ignition energy decreases with increasing $\rho_0$ under the law $E_{\mathrm{ig}}\sim \rho_0^{-2}$ in accordance with the known theoretical estimate \citep{Tabak-Hammer-Glinsky-Kruer-Wilks-Woodworth-Campbell-Perry-Mason-1994}. For $\rho_0 = 100 \rho_s \approx 22$~g/cm$^3$, the ignition energy $E_{\mathrm{ig}} = 160$~kJ.

\section{Conclusions}\label{conclusions}

Computational experiments with two proton beams of the same energy and different intensities show that the beam intensity required for the ignition increases with the initial fuel density in accordance with a known formula, obtained by results of two-dimensional calculations.

If the target is ignited, a detonation wave, which is adjacent to the front of the rarefaction wave, occurs. The formation of the detonation wave precedes a short time interval with a shock wave that goes ahead of a slow combustion wave. When the combustion front overtakes the shock wave, the latter transforms into the detonation wave.

There is no ignition if the distance between the shock front and the region of the fuel heating by protons is too large at the end of irradiation. Over time, the rarefaction wave overtakes the shock front and reduces the intensity of the shock wave.

For small values of $R_{\alpha}$, limiting trajectories $\alpha$-particles, and small target thickness ($H\rho_0 \approx 1$~g/cm$^2$), the target ignition is possible after the shock wave reflection from the symmetry plane. With increasing $H$, starting from a certain value, the target ceases to ignite due to the aforementioned decrease in the intensity of the shock wave that is overtaken by the rarefaction wave.

Upon reflection of the detonation wave from the symmetry plane, the flow with the linear velocity profile along the spatial variable $x$ and with a weak dependence of the thermodynamic functions upon $x$, close to one of the representatives of a certain family of solutions of hydrodynamic equations, occurs between this plane and the reflected detonation wave. Combustion efficiency at this stage is largely due to the spatial homogeneity of the burn-up rate (the product of the DT reaction rate and the triton number density) whose integral along the trajectory of a Lagrangian particle defines the local burn-up factor.

Solution of the equations of hydrodynamics with a linear velocity profile and constant in $x$ values of thermodynamic functions is found analytically up to an arbitrary constant. The possibility to choose this constant so that the solution describes with good accuracy the change in the density with time after reflection of the detonation wave is shown.

For the parameter $H\rho_0 \approx 1$~g/cm$^2$, the main contribution to the final value of the burn-up factor $B \approx 0.3$ is given by the expansion stage after reflection of the detonation wave. For $H \rho_0 \approx 5$~g/cm$^2$, the burn-up factor is increased by the expansion stage from 0.4 to 0.7. At $H\rho_0 \approx 1$~g/cm$^2$, the gain with respect to the energy of neutrons $G \approx 200$. At $H\rho_0 \approx 5$g/cm$^2$, the burn-up factor increases by more than a factor of 2, which, together with the increase in the fuel mass 5 times, gives the gain $G>2000$. The above values of the burn-up factor and the gain should be considered as the maximum possible values for the real cylindrical target of the same initial density and length.

The quasi-1D model, limiting trajectories of $\alpha$-particles by a cylinder of a given radius, reproduces known theoretical dependence of the ignition energy $E_{\mathrm{ig}} \sim \rho_0^{-2}$. For $\rho_0 = 100\rho_s \approx 22$~g/cm$^3$, the quasi-1D model gives $E_{\mathrm{ig}} = 160$~kJ.

\section{Acknowledgments}
The work is supported by grants of the Russian Foundation for Basic Research (12-01-00130 and 14-08-00967) and the President of the Russian Federation (NSh-6614.2014.2), as well as by the programs of scientific investigations of the Russian Academy of Sciences (program 3 of the Division of Mathematics RAS and program 2P of the Presidium RAS).

\appendix
\setcounter{section}{1}
\section*{Appendix: Track method of plasma heating by charged products of a thermonuclear reaction and kinetic Fokker--Plank equation}

We limit our consideration by plane 1D flows for which all of the plasma parameters depend upon only one space variable $x$. Without loss of generality, the domain $0 \leqslant x \leqslant 1$ can be considered. The charged products of a thermonuclear reaction will be referred to as particles. The deceleration $a(x,v) < 0$, the number of the reaction events per unit time and unit volume $F(x)$ and the velocity of particle thermalization $v_{\mathrm{th}}(x)$ are supposed to be given.

To simplify calculations, at first we consider the problem without particles entering the domain. At the end of the section, we present the modification of the method for the condition of completely elastic reflection of particles from one of the boundaries, which simulates the symmetry condition at this boundary.

Consider the track method \citep{Brueckner-Jorna-1973, Duderstadt-Moses-1982} based on simple physical reasons applied to a discrete media. $F(x_0)$ particles with the same magnitude of velocity $v_0$ are produced in unit volume per unit time at the point $x_0$. The particles move in straight rays in all directions, which of them is given by the cosine $\mu$ of the angle between the ray direction and the axis $x$, $-1 \leqslant \mu \leqslant 1$. Consider the particle velocity $v(\xi,x_0,\mu)$ determined by the Cauchy problem
\begin{equation} \label{PartMov}
v\frac{\partial v}{\partial \xi} = a(x,v), \quad x=x_0+\xi\mu, \quad v(0,x_0,\mu)=v_0,
\end{equation}
where $\xi$ is the coordinate along the ray. Since $a < 0$, the particle velocity decreases. The maximal value of $\xi$ for given $x_0$ and $\mu$ is determined by the conditions
\begin{equation} \label{KsiLimit}
0\leqslant x_0+\xi_{\max}\mu \leqslant 1, \quad v(\xi_{\max},x_0,\mu)= v_{\mathrm{th}}(x_0+\xi_{\max}\mu).
\end{equation}

Let the grid $x_l$, $l = 1$, 2, \dots, $N+1$, $x_1 = 0$, $x_{N+1} = 1$ divides the coordinate $x$ into the cells of the size $\Delta x_l = x_{l+1} - x_l$, $l = 1$, 2, \dots, $N$. For definiteness, assume the plasma parameters are defined at the centers of the cells $\bar x_l = 0.5(x_l+x_{l+1})$.

Number of particles produced in the cell $j$ with unit cross-section area per unit time, $\sigma_j = F(\bar x_j) \Delta x_j$. The particles are divided into $M$ angular groups with the constant step of $\mu$, $\Delta\mu = 2/M$, basing on the isotropic angular distribution of particles ($\sigma_j\Delta\mu/2$ particles in the group). All particles of one group $m$ produced in the cell $j$ move along the straight ray with $\mu = \mu_m = -1 + (m - 0.5)\Delta\mu$ and are decelerated in compliance with Eq. \eref{PartMov} with $x_0 = \bar x_j$. We restrict our consideration by the case of even values of $M$, for which $\mu_m \neq 0$ at any $m$.

Let the particles of a certain angular group produced in the cell $j$ traverse the cell $k$ of dimension $\Delta x_k$. If $k = j$, the boundary condition for Eq. \eref{PartMov} is given at the cell center, $v = v_0$. Otherwise, the boundary condition is given at one of the points $x_k$ or $x_{k+1}$ with a smaller value of $\xi$, $v = v_{0b}$, where the velocity $v_{0b}$ has been obtained from computation of the previous cell. In the most simple variant of the method, which is considered in the present paper, the function $a(x,v)$ within the cell supposes to be of constant value $a_k=a(\bar x_k, v_b)$, where $v_b = v_{0b}$ at $k \neq j$ and $v_b = v_0$ at $k = j$. The velocity $v_e$ at the point with the greater value of $\xi$ is determined by the Cauchy problem (\ref{PartMov}) and the thermalization condition $v_e \geqslant v_{\mathrm{th}}^k = v_{\mathrm{th}}(\bar x_k)$
\begin{equation*} 
\tilde v_e^2=v_b^2+2 a_k \Delta\xi, \quad v_e=\max{(v_{\mathrm{th}}^k,\tilde v_e)},
\end{equation*}
where $\Delta\xi = \Delta\xi_k = \Delta x_k / |\mu_m|$ at $k \neq j$ and $\Delta\xi = \Delta\xi_k/2$ at $k = j$. If particles are thermalized ($\tilde v_e \leqslant v_{\mathrm{th}}^k$), computation of the given angular group is terminated.

The energy transferred by particles of the angular group $m$ produced in the cell $j$ to unit volume of the cell $k$ in unit time
\begin{equation}
\label{psi-kjm}
\psi_{kjm}=\frac{m_p\sigma_j\Delta\mu}{2\Delta x_k}\frac{v_b^2-v_e^2}{2}=
\frac{m_p\sigma_j \Delta\mu}{2|\mu_m|}
\left\{
\begin{aligned}
&\min\bigg(-a_k,\frac{v_{0b}^2-(v_{\mathrm{th}}^k)^2}{2\Delta\xi_k}\bigg), & k \neq j, \\
&\min\bigg(-\frac{a_k}{2},\frac{v_0^2-(v_{\mathrm{th}}^k)^2}{2\Delta\xi_k}\bigg), & k=j, \end{aligned}
\right.
\end{equation}
where $m_p$ is the particle mass. The total energy transferred to unit volume of the cell $k$ in unit time
\begin{equation} \label{TotEnDis}
w_k=\sum_{m=1}^M \sum_{j=k}^{k+\Delta k} \psi_{kjm},
\end{equation}
where $k + \Delta k$ is the number of the cell in which the particles of group $m$, those are thermalized in the cell $k$, were produced. The sign of $\Delta k$ depends of the sign of $\mu_m$: $\Delta k \leqslant 0$ at $\mu_m > 0$ and $\Delta k \geqslant 0$ at $\mu_m < 0$. The evident restriction on $\Delta k$ is $1 \leqslant k + \Delta k \leqslant N$. To compute the sum \eref{TotEnDis}, the terms $\psi_{kjm}$ are computed step-by-step for all of the cells $1 \leqslant j \leqslant N$ and the angular groups $1 \leqslant m \leqslant M$.

To study connection between the track method and the kinetic steady-state equation in Fokker--Plank approximation \citep{Guskov-Rozanov-1982-eng}, it is necessary to formulate the track method in terms of the mathematical physics rather than algebraic equations, i.e. to find the limit of $w_k$ at $h = \max(\Delta x_l)$, $\Delta\mu\to 0$.

Let, to begin with $h \to 0$. Then, the number of terms in the internal sum of Eq. \eref{TotEnDis} tends to infinity. Therefore, one can reject the first term with $j = k$ and the last term with $j = k+\Delta k$. As a result, the internal sum in Eq. \eref{TotEnDis} takes the form
\begin{equation} \label{Psi-km}
\Psi_{km}=-\frac{m_p\Delta\mu}{2|\mu_m|}\sum_{j=k+\mathrm{sign}(\Delta k)}^{k+\Delta k-\mathrm{sign}(\Delta k)}
a(\bar x_k,v_{b0}) F(\bar x_j) \Delta x_j.
\end{equation}

Denote $x=\lim_{h \to 0}\bar x_k$, $x_0=\lim_{h \to 0}\bar x_j$. As long as $\lim_{h \to 0} x_k = \lim_{h \to 0} x_{k+1} = x$, $\lim_{h \to 0} v_{b0} = v(\xi,x_0,\mu_m)$, $\xi = (x - x_0)/\mu_m$, where $v(\xi,x_0,\mu_m)$ is the solution of the Cauchy problem \eref{PartMov}. As a result, obtain
\begin{gather} \label{LImPsi-km}
\lim_{h \to 0}\Psi_{km} = \Psi(x,\mu_m)= \cr =\frac{m_p\Delta\mu}{2\mu_m}\int\limits_x^{x+\Delta x} a\big(v(\xi,x_0,\mu_m),x\big) F(x_0) \mathrm{d}x_0, \end{gather}
where $\Delta x$ is determined by thermalization condition at the point $x$ for particles of group $m$ produced at the point $x+\Delta x$:
\begin{equation} \label{DxCond}
v(-\Delta x/\mu_m, x+\Delta x, \mu_m) = v_{\mathrm{th}}(x).
\end{equation}

Disappearance of the modulus sign for $\mu_m$ under conversion from the sum \eref{Psi-km} to the integral \eref{LImPsi-km} is due to that the sum \eref{Psi-km} conserves its sign at changing the sign of $\Delta k$, while the integral \eref{LImPsi-km} changes its sign at changing the sign of $\Delta x$.

At $\mu_m \to 0$, the function $\Psi(x,\mu_m)$ remains finite though $\mu_m$ is in the denominator of the function. We limit our proof by the simplest case $a = \mathrm{const}$. Then the condition \eref{DxCond} and the Cauchy problem \eref{PartMov} give $\Delta x = \mu_m\big(v_0^2 - v_{\mathrm{th}}^2(x)\big)/2a$, from which 
$$
\Psi(x,0) = \frac{m_p\Delta\mu F(x)\big(v_0^2 - v_{\mathrm{th}}^2(x)\big)}{4}.
$$

Finally, the required limit, which is the rate of energy transfer to unit volume of plasma at the point $x$, is as follows:
\begin{gather} \label{tr-meth}
W(x)=\lim_{h,\Delta\mu \to 0} w_k = \cr = \frac{m_p}{2} \int\limits_{-1}^1 \int\limits_x^{x+\Delta x} \frac{a\big(x,v(\xi,x_0,\mu)\big) F(x_0)}{\mu} \mathrm{d}x_0 \mathrm{d}\mu,
\end{gather}
where $v(\xi,x_0,\mu)$ is the solution of the Cauchy problem \eref{PartMov}, $\Delta x$ is determined by the condition \eref{DxCond}.

Now we consider the kinetic steady-state equation for the distribution function $f(x,v,\mu)$ in the Fokker--Plank approximation. If all of the produced particles have the same initial velocity $v_0$ and the isotropic angular distribution, as well as the diffusion of the function $f(x,v,\mu)$ in the velocity space can be ignored, the considered equation reduces to the following Cauchy problem for the kinetic homogeneous equation
\citep{Guskov-Rozanov-1982-eng}
\begin{equation} \label{ini-kin-eqA}
\mu v\frac{\partial f}{\partial x} + \frac{\partial af}{\partial v} =0, \quad 
v_{\mathrm{th}}(x)\leqslant v \leqslant v_0, \quad 
f(x,v_0,\mu)=-\frac{F(x,\mu)}{2a(x,v_0)}.
\end{equation}

The rate of energy transfer to unit volume of plasma at the point $x$ has the form
\begin{equation} \label{ini-en}
W_{\mathrm{FP}}(x) = \int\limits_{-1}^1 \Psi_{\mathrm{FP}}(x,\mu)\mathrm{d}\mu, \quad 
\Psi_{\mathrm{FP}}(x,\mu)= -m_p \int\limits_{v_{\mathrm{th}}(x)}^{v_0} f(x,v,\mu) a(x,v) v\mathrm{d}v.
\end{equation}

Instead of the velocity $v$, we introduce the variable $\eta = v^2/2$, setting $f = f(x,\eta,\mu)$, $a = a(x,\eta)$, $\eta_0 = v_0^2/2$, $\eta_{\mathrm{th}}(x) = v_{\mathrm{th}}(x)^2/2$. Then the function $\Psi_{\mathrm{FP}}$ from Eq. \eref{ini-en} takes the form
\begin{equation} \label{en-mu}
\Psi_{\mathrm{FP}}(x,\mu)= -m_p \int\limits_{\eta_{\mathrm{th}}(x)}^{\eta_0} f(x,\eta,\mu) a(x,\eta) \mathrm{d}\eta.
\end{equation}

Using the change of variable $x=x_0 + \xi\mu$, we introduce the new independent variable $\xi$ and the parameter $x_0$, which sense will be specified below. Equation \eref{ini-kin-eqA} takes the form
\begin{equation} \label{kin-eq}
\frac{\partial f}{\partial \xi} + a\frac{\partial f}{\partial \eta} + fa_{\eta}=0, \quad f=f(x_0+\xi\mu,\eta,\mu).
\end{equation}
Its characteristics are determined by the differential equation $\mathrm{d}\eta = a\mathrm{d}\xi$ and have the form of the one-parameter family
\begin{equation} \label{charact}
\eta(\xi,x_0,\mu)=\eta_0+\int\limits_0^{\xi}a\big(x_0+\xi'\mu,\eta(\xi',x_0,\mu)\big)\mathrm{d}\xi'.
\end{equation}
where the parameter $x_0$ defines the characteristic that pass through the point $(x_0, \eta_0)$. Note, that the above family rewritten in another notations coincides with the family of solutions of the Cauchy problem \eref{PartMov}.

Along the characteristics \eref{charact}, equation \eref{kin-eq} and the boundary condition from the Cauchy problem \eref{ini-kin-eqA} take the form
\begin{equation*} 
\frac{\mathrm{d}f}{\mathrm{d}\xi} + fa_{\eta}=0, \quad
f=f\big(x_0+\xi\mu,\eta(\xi,x_0,\mu),\mu\big), \quad
f(x_0,\eta_0,\mu)=-\frac{F(x_0)}{2a(x_0,\eta_0)},
\end{equation*}
that gives
\begin{gather} \label{solut-kin}
\begin{split}
f\big(x_0+\xi\mu,\eta(\xi,x_0,\mu),\mu\big)=-\frac{F(x_0)\chi(\xi,x_0,\mu)}{2a(x_0,\eta_0)}, \\
\chi(\xi,x_0,\mu)=\exp\Bigg(-\int\limits_0^{\xi} a_{\eta}\big(x_0+\xi'\mu,\eta(\xi',x_0,\mu)\big)\mathrm{d}\xi'\Bigg).
\end{split}
\end{gather}


In Eq. \eref{en-mu}, the integration over $\eta$ is replaced by the integration over $x_0$ connected with $\eta$ by Eq. \eref{charact} and by the relation $x = x_0 + \xi\mu = \mathrm{const}$. To simplify calculations, the relation between the differentials $\mathrm{d}\eta$ and $\mathrm{d}x_0$ can be found by going from the variable $\xi$ to the variable $x$ in Eq. \eref{charact}, taking the differential of Eq. \eref{charact} and setting $\mathrm{d}x = 0$. As a result, we obtain
\begin{equation} \label{dh-dx0}
\mathrm{d}\eta = -\frac{a(x_0,\eta_0)}{\mu}\mathrm{d}x_0.
\end{equation}

The limits of integration $\eta_0$ and $\eta_{\mathrm{th}}(x)$ are changed to $x$ and $x + \Delta x$ respectively, where $\Delta x$ is determined by the condition $\eta( - \Delta x/\mu,x + \Delta x,\mu)=\eta_{\mathrm{th}}(x)$ similar to the condition \eref{DxCond} in the track method. By replacing the limits in Eq. \eref{en-mu} and using Eqs. \eref{solut-kin} and \eref{dh-dx0}, we obtain
\begin{equation}
\label{en-mu-dx0}
\Psi_{\mathrm{FP}}(x,\mu)= \frac{m_p}{2}\int\limits_x^{x+\Delta x} \frac{F(x_0)\chi(\xi,x_0,\mu) a\big(x,\eta(\xi,x_0,\mu)\big)}{\mu}\mathrm{d}x_0. 
\end{equation}

One can see that the inverse change of variable $\eta$ by variable $v$ in Eq. \eref{en-mu-dx0} and the substitution \eref{en-mu-dx0} into \eref{ini-en} give the function $W_{\mathrm{FP}}(x)$ differed from the function \eref{tr-meth} in the track method by only the multiplier $\chi(\xi,x_0,\mu)$, which is defined by Eq. \eref{solut-kin}. Therefore, the track method described at the beginning of the section for finite numbers of the grid cells on $x$ and the angular groups on $\mu$ can be used as the numerical method for the problem \eref{ini-kin-eqA}, \eref{ini-en} with the additional computation of the function $\chi(\xi,x_0,\mu)$ along the every trajectory of particles.

To compute the rate of the energy transfer to unit of plasma volume separately for electrons and ions, the decelerations of particles by plasma components $a_e(x,v)$ or $a_i(x,v)$ should be used in Eq. \eref{psi-kjm} instead of $a(x,v)$.

If one of the boundaries, $x = 0$, for example, is the symmetry plane, the condition of symmetrical reflection of the particle trajectory from the boundary at $\mu < 0$
$$
x=\begin{cases}
x_0+\xi\mu,& \xi\leq \xi_* = -x_0/\mu, \\
-(\xi-\xi_*)\mu,& \xi>\xi_*,
\end{cases}
$$
is added to the problem \eref{PartMov}, and the condition $x_0 + \xi_{\max}\mu \geqslant 0$, that determines maximal $\xi$ value, is removed from the conditions \eref{KsiLimit}.

In the case of the quasi-1D model, the condition \eref{quasi1D} is added to the conditions \eref{KsiLimit}.



\hyphenation{Post-Script Sprin-ger Milyav-s-kii Khish-chen-ko Che-l-ya-b-insk
  But-ter-worth}

\end{document}